\documentclass[pre,aps,twocolumn]{revtex4}
\usepackage{bm}
\usepackage{graphicx}
\usepackage{amsmath}
\usepackage{amssymb}
\begin{document} 
\title{Parametric perspective on highly excited states: case studies
of CHBrClF and C$_{2}$H$_{2}$} 
\author{Aravindan Semparithi and Srihari Keshavamurthy} 
\affiliation{Department of Chemistry, Indian Institute
of Technology, Kanpur, India 208 016}
\begin{abstract}
Considerable insights can be obtained
regarding the nature of highly excited states by computing the
eigenbasis expectation values of the resonance operators associated with
an effective spectroscopic Hamiltonian. The expectation values
are related
to the parametric derivative
of the eigenvalues with respect to specific resonance strengths
{\it i.e.,} level velocities. 
Sensitivity of the level velocities to the existence of
closed orbits in the underlying classical phase space 
provides for a dimensionality independent route to a dynamical
assignment of the states. In this letter, taking 
CHBrClF (polyad $P=5$)
and the $[16,0]^{g+}$ bend polyad of C$_{2}$H$_{2}$ as examples, we
show that the level velocities can signal the birth of new modes
and highlight sequences of localized eigenstates.
\end{abstract}
\maketitle
\nopagebreak

\section{Introduction}

Dynamical assignment of the highly excited eigenstates of a polyatomic
molecule is a topic of significant interest to the chemical
physics community\cite{rev1,rev2,rev3,rev4}. 
Insights into the nature of
the excited states provides for a better understanding of the
phenomenon of intramolecular vibrational energy redistribution (IVR)
occuring in the molecule\cite{rev1,rev3,rev4}. 
Assignment of the low energy vibrational states
in terms of the usual normal mode quantum numbers is relatively
straightforward. However with increasing energies the normal mode
quantum numbers are no longer conserved due to perturbations
that strongly mix the zeroth order modes and the 
molecular vibrational Hamiltonian\cite{rev4}
takes the form:
\begin{eqnarray}
\widehat{H} &=& \sum_{i=1}^{N} 
\nu_{i} \hat{v}_{i} + \sum_{i<j=1}^{N} x_{ij} \hat{v}_{i}
\hat{v}_{j} +\ldots + \sum_{k}^{} \tau_{k} \widehat{V}_{k}(\hat{\bf a}_{k},
\hat{\bf a}^{\dagger}_{k}) \nonumber \\
&\equiv& \widehat{H}_{0}(\hat{\bf v}) + 
\sum_{k}^{} \tau_{k} \widehat{V}_{k}(\hat{\bf a}_{k},\hat{\bf a}^{\dagger}_{k})
\end{eqnarray}
wherein $H_{0}$ represents the Dunham expansion, $x_{ij}$ are the
anharmonic constants and $\widehat{V}_{k}$ are the various perturbations.
The operators $\hat{a}_{k}$ and $\hat{a}^{\dagger}_{k}$ represent the
annhilation and creation operators for the $k^{\rm th}$ mode
respectively.
The normal mode quantum numbers 
$v_{i} = \hat{a}^{\dagger}_{i} \hat{a}_{i}$, sufficient for assigning low
energy eigenstates, do not commute with $\widehat{V}_{k}$ and hence cannot
be used to assign the eigenstates of $\widehat{H}$.
The above Hamiltonian is usually generated 
from a fit to the experimental
spectrum in the absence of a global {\it ab initio} potential energy
surface. Even in the case when an accurate potential energy surface
is at hand it is possible, and useful, to generate a Hamiltonian
of the above form 
using the canonical Van-Vleck perturbation theory\cite{cvpt1,cvpt2}.
The advantages of using Hamiltonians of the above form stems from 
the fact that the classical limit Hamiltonian 
\begin{equation}
H({\bf I},{\bm \theta}) = H_{0}({\bf I})+2 \sum_{k} \tau_{k}
f_{k}({\bf I}) \cos({\bf m}_{k} \cdot {\bm \theta})
\end{equation}
is easily obtained
via the correspondence
$\hat{\bf a} \leftrightarrow \sqrt{\bf I}
e^{-i {\bm \theta}}$. 
The $f$-dimensional vector ${\bf m}_{k}$ has integer
components $(r_{1},r_{2},\ldots,r_{f})$ and 
$({\bf I},{\bm \theta})$ are the action-angle variables\cite{aav}
corresponding to $H_{0}$. 
The order of a resonance\cite{aav} is defined 
as $O_{k} \equiv \sum_{j=1}^{f} |r_{j}|$.
The perturbations $V_{k}({\bf I},{\bm \theta}) \equiv 
f_{k}({\bf I}) \cos({\bf m}_{k} \cdot {\bm \theta})$ are called resonant
because the condition ${\bf m}_{k} \cdot \Omega({\bf I})=0$ implies
a specific commensurability or locking between the unperturbed
frequencies $\Omega({\bf I}) \equiv \partial H_{0}({\bf I})/\partial {\bf I}$.
Such resonances are responsible for energy flow
through the molecule and lead to breakdown of the zeroth order, low energy
quantum numbers. 
Classical dynamics of resonant Hamiltonians
is historically very rich and have
been studied in great detail\cite{aav}. 
In particular a detailed understanding
of the classical dynamics of  
$H({\bf I},{\bm \theta})$ is essential for any dynamical assignment
of the quantum eigenstates of $\widehat{H}$. 
Furthermore the 
effective Hamiltonian approach utilizes the state space perspective
which offers considerable advantages towards understanding IVR\cite{rev4}.

Clearly any 
assignment of the eigenstates of $\widehat{H}$
formally requires the existence of a sufficiently large set 
of good quantum numbers. Such a set 
does not exist in general and hence
assignment implies the
existence of at least approximate or quasi quantum numbers. 
By necessity such approximate quantum numbers are conserved
for a certain time period dictated by the dynamics of the system.
Support for the notion of quasi quantum numbers partially comes from
the experimental observation\cite{rev4} of hierarchical IVR in molecules and
the fact that most molecules are not ergodic even at fairly high
energies\cite{leto}.
In this sense deciphering the quantum numbers, exact or approximate, invariably
implies the knowledge of the underlying dynamics.
Consequently one speaks of a dynamical assignment of the eigenstates
wherein some or all of the quantum numbers arise by focusing
on important quantum and or classical dynamical structures. 
A consistent picture that is emerging from 
many studies is that
the complicated spectral splittings and patterns at high energies 
can be `unzipped' to some extent using the underlying classical dynamics. 
More specifically, {\em the spectra are unzipped by recognizing eigenstates
showing similar localization characteristics about important classical
invariant structures in the underlying phase space}.

Studies based on classical-quantum correspondence have
been successfully applied 
to systems with two 
coupled modes\cite{kell,joy,jac,tay1,tay2,field,ksgse1,kspccp}
but are yet to be extended to 
systems with three or more coupled modes {\it i.e.,} 
multimode systems\cite{commcdbr}. 
The technical and conceptual difficulties associated with a straightforward
generalization of the two mode
techniques to multimode systems are well understood\cite{ksgse1}.
Circumventing the technical difficulty, in our opinion, requires 
utilizing quantum objects which are sensitive to the underlying classical
mechanics but do not rely on visualizing the phase space and eigenstates.
Considerable work has been done in this direction and a common theme
underlying such approaches is the analysis of the eigenvalues and their
variation with Hamiltonian 
parameters\cite{ram,rose,wu,kay,ksjpca,child}. 
For instance the nature of an eigenstate $|\alpha \rangle$ of $\widehat{H}$
has been studied using the 
methods of diabatic correlations\cite{rose,wu} revealing
the existence of formal quantum numbers. The basic quantity in these studies
is the variation of the eigenvalue $E_{\alpha}$ with a specific coupling
strength {\it i.e.,} $\partial E_{\alpha}/\partial \tau_{k}$.
Remarkable correlation of the level variations
to the phase space nature of the eigenstates 
had been noted by Weissman and Jortner in
the context of the Henon-Heiles system\cite{weissjort}. 
Support for the 
correlation was provided recently\cite{ksjcp} 
from a semiclassical viewpoint and
it was suggested that the parametric variations were sensitive to
the various bifurcations occuring in the classical phase space. 
However strong support for the observed 
correlations have existed\cite{eck}
in the literature in terms of the classical-quantum correspondence
of quantum expectation values {\it i.e.,} diagonal matrix elements.

In order to elucidate the connections we note that 
the Hellman-Feynman theorem
\begin{eqnarray}
V_{k}^{\alpha \alpha} &\equiv& \langle \alpha|\widehat{V}_{k}|\alpha \rangle 
=\frac{\partial E_{\alpha}}{\partial \tau_{k}} \\
&=& 2 \sum_{P_{k}} V_{k}^{\alpha \alpha}({\bm \tau};P_{k})
\end{eqnarray}
suggests the diagonal matrix element of the perturbation as the
fundamental object. 
$P_{k}$ represents the polyad, associated with the
resonance $\widehat{V}_{k}$, whose constancy is destroyed in the
presence of other independent resonant perturbations.
Consequently\cite{ksjcp,kspccp}  
dominance of $V_{k}^{\alpha \alpha}({\bm \tau};P_{k})$ 
at a single $P_{k}$ implies a highly localized $|\alpha \rangle$
in the state space with $P_{k}$ being an approximate quantum number.
Quantum mechanically if  
$|{V}_{k}^{\alpha \alpha}|$ is large then it is
expected that the perturbation $\widehat{V}_{k}$ plays a role
in determining the nature of $|\alpha \rangle$. 
On the other hand a semiclassical analysis of
the expectation value provides valuable information on the phase space
nature of $|\alpha \rangle$. This can be seen by considering the quantity
$\rho_{k}(E) \equiv 
\sum_{\alpha} V_{k}^{\alpha \alpha} \delta(E-E_{\alpha})$ 
which is
the expectation values of $\widehat{V}_{k}$
weighted by the density of states. 
Such quantities lend themselves to an elegant semiclassical interpretation
in terms of the classical closed orbits in the phase space.
This is hardly surprising given that the genesis of
the Gutzwiller periodic orbit quantization\cite{gutz} idea was from
a semiclassical analysis of the quantum density of states
$\rho = \sum_{\alpha} \delta(E-E_{\alpha})$.
As the method of semiclassical analysis of $\rho_{k}(E)$ is
well established in the literature and our intention in this work
is not to semiclassically evaluate $V_{k}^{\alpha \alpha}$ we will
highlight the salient features.
In general $\rho_{k}$ 
has a smooth and oscillating part with
the smooth part being independent of
the nature of the classical dynamics. The oscillating part
is sensitive to the nature of the dynamics
and can be written down in terms of a sum over the closed orbits in the
phase space. If the phase space is chaotic then the 
closed orbits are the various periodic orbits whereas for a regular
phase space the closed orbits are the rational tori. In either case
it can be shown\cite{eck} that the oscillating part depends inversely on the
determinant of the stability matrix $M$ 
of the closed orbit and directly on the quantity
\begin{equation}
V_{kp} = \frac{1}{T_{p}} \int_{0}^{T_{p}} dt V_{k}({\bm \theta}(t),{\bf I}(t))
\end{equation}
representing the average of the resonant term over one period $T_{p}$
of the closed orbit.

We emphasise the dependence of
$\rho_{k}$ on the quantities $M$ and $V_{kp}$ for two reasons.
Firstly, $V_{kp}$ clearly underscores the important role played by the
classical analog of $\widehat{V}_{k}$.
Further, performing a standard canonical
transformation $({\bf I},{\bm \theta}) \rightarrow (J,\psi,{\bf K},{\bm \chi})$
with $(J,\psi)$ being the slow angle and action variables specific
to the resonant term $\cos({\bf m}_{k} \cdot {\bm \theta})$ we obtain
\begin{equation}
V_{kp} = \frac{1}{T_{p}} \int_{0}^{T_{p}} dt f_{k}(J,{\bf K}) \cos \psi
\end{equation}
Now for a closed orbit in the phase space corresponding to 
${\bf m}_{k} \cdot {\bm \theta} = 0$ the angles ${\bm \chi}$ are 
fast and can be averaged resulting in
the actions ${\bf K}(t) \approx {\bf K}(0)$. The fixed points
for the averaged system are then determined by $(\dot{J}_{p},\dot{\psi}_{p})=
(0,0)$ and correspond to the closed orbit in the full phase space.
Within this averaged viewpoint 
$V_{kp} \approx \pm f_{k}(J_{p},{\bf K})$ with the signs
coming from $\psi_{p}=0,\pm \pi$. Thus it is expected that
a maximum in $|V_{k}^{\alpha \alpha}|$ comes from the localization of
the eigenstate $|\alpha \rangle$ due to closed orbits associated with
$\widehat{V}_{k}$. Evidently states influenced
by a particular closed orbit can be classified into a group and
identified by patterns in the $V_{k}^{\alpha \alpha}$ ``spectrum''.
In mixed phase space regimes, generic to molecular systems, periodic
orbits with varied stabilities can exist and influence the dynamics.
In the following sections we show the expectation values which are scaled
to unit variance and zero centered. The scaling is performed to remove
the dependence\cite{ksjpca,ksjcp} 
of $V_{k}^{\alpha \alpha} \propto P_{k}^{O_{k}/2}$ 
on the approximate polyad
$P_{k}$ arising from the localization of the eigenstate $|\alpha \rangle$  
due to $\widehat{V}_{k}$. 

Secondly, bifurcations in the phase space
are signalled by the vanishing\cite{aav} 
of the determinant ${\rm det}|(M-1)|$. 
In general bifurcations imply birth of new orbits and/or death of
old orbits. In the molecular context\cite{kell} such bifurcations manifest
themselves as destruction of the old modes and creation of the new modes which
influence the eigenstates. For instance a normal to local transition with
varying energy arises due to bifurcations in the classical phase space.
Thus it is natural to expect that the $V_{k}^{\alpha \alpha}$ ``spectrum''
will exhibit the effects of the various bifurcations. 

Recent work from our group has demonstrated the usefulness of the above
method for understanding the highly excited states of a 
model system\cite{ksjcp} and
the DCO radical\cite{kspccp}. 
In this letter
two molecules, CHBrClF and C$_{2}$H$_{2}$,
are chosen as further examples to illustrate the method.
Specifically, the $N=5$ polyad of CHBrClF and
the $[16,0]^{g+}$ bending polyad 
of C$_{2}$H$_{2}$ are investigated. 
In section II
the highly excited states of CHBrClF are analyzed and compared to a recent
work\cite{tay2}. 
We find some discrepancy in the dynamical assignments provided
earlier and by the level velocity method. In section III the pure bending
states of C$_{2}$H$_{2}$ are investigated and it is shown that the level
velocity approach is capable of identifying the new class of
local bending and counter-rotation states. Comparing 
to a recent work\cite{tay3} we show that
the level velocity approach is successful in identifying important
eigenstate sequences at such high levels of excitation. Section IV concludes.

\section{Dynamical nature of CHBrClF eigenstates in Polyad $N=5$}

An effective Hamiltonian for CHBrClF was proposed
by Beil {\it et al.} on the basis of their detailed study of
the rovibrational spectra\cite{beil}. 
The CH overtone spectrum implicated multiple
Fermi resonances involving the pure CH stretch $(s)$ 
and the two $(a,b)$ CH
bending vibrations. The spectroscopic Hamiltonian has the
form:
\begin{equation}
\widehat{H} = \widehat{H}_{0} + k_{saa} \widehat{V}_{saa} 
+ k_{sbb}
\widehat{V}_{sbb} +  k_{sab} \widehat{V}_{sab} 
+ \gamma \widehat{V}_{aabb}
\end{equation}
The zeroth-order anharmonic part is diagonal in the 
number $(v_{s},v_{a},v_{b})$ basis
\begin{equation}
H_{0}=\sum_{j}^{sab} \nu_{j} v_{j} + \sum_{j}^{sab}\sum_{k}^{sab}
x_{jk} v_{j} v_{k}
\end{equation}
with the $\nu_{j}$ and $x_{jk}$ being the 
harmonic frequencies and anharmonicities of the three modes respectively.
The perturbations $\widehat{V}$ are off-diagonal 
in the $(v_{s},v_{a},v_{b})$ basis and represent resonant coupling
of the modes. 
The first three perturbation
terms in the Hamiltonian are Fermi resonances between the CH stretch
and the various bend modes whereas the last perturbation is a 
Darling-Dennison resonance between the two bend modes. 
The structure of the effective Hamiltonian
implies the existence of a constant of the motion $N=v_{s}+(v_{a}+v_{b})/2$
called as the polyad number. 
For the various
parameter values and form of the resonant operators
we refer the reader to the earlier works\cite{beil,tay2}.
However, we note that the bending mode anharmonicities
$x_{aa}, x_{bb}, x_{ab}$ are small and the resonant coupling strengths
are rather large. For instance, $x_{aa} \sim -6$ cm$^{-1}$ and
the stretching mode anharmonicity $x_{ss} \sim -65$ cm$^{-1}$ whereas
$k_{sbb} \sim 113$ cm$^{-1}$. Such large resonant strengths combined with
reduced bend anharmonic constants imply that any analysis solely
based on $H_{0}$ would be inadequate. 

We also report the inverse participation ratios (IPR) of
the eigenstates in various basis. The IPR $L_{\alpha}$ of an
eigenstate $|\alpha \rangle$ in a basis $|b\rangle$ is given
by $L_{\alpha}=\sum_{b}|\langle b|\alpha \rangle|^{4}$. $L_{\alpha}$
is a measure of the extent of delocalization of an eigenstate in
a specific basis. A high $L_{\alpha}$ indicates localization in
the state space whereas for a completely delocalized state 
$L_{\alpha}=1/N_{s}$ with $N_{s}$ being the total number of states.

\begin{figure}
\includegraphics*[width=3.5in,height=3.5in]{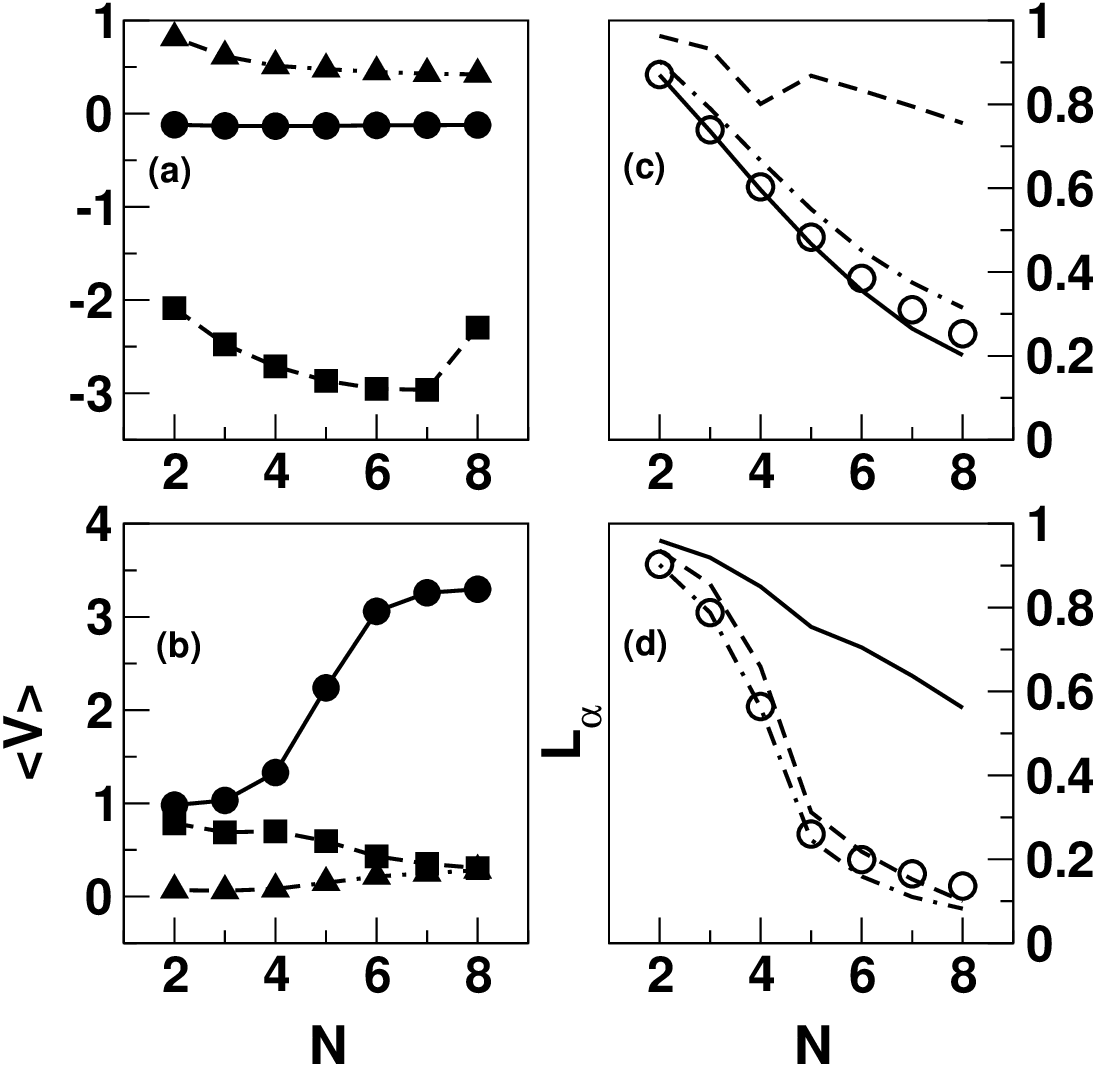}
\caption{Resonance expectation values as a function
of the polyad $N=v_{s}+(v_{a}+v_{b})/2$ for
(a) lowest and (b) highest energy states in a polyad for CHBrClF.
The circles, squares and triangles represent the
expectation values of $\widehat{V}_{saa}$, $\widehat{V}_{sbb}$,
and the $\widehat{V}_{aabb}$ respectively.
The IPRs for the two sets of states are shown in (c) and (d).
Open circles denote IPRs in the zeroth order basis and
the solid line, dashed line, dash-dotted lines represent
IPRs in the $H_{0}+V_{saa}$, $H_{0}+V_{sbb}$, $H_{0}+V_{aabb}$
basis respectively.}
\label{fig1}
\end{figure}

In this work we focus on the $N_{s}=36$ eigenstates belonging to the polyad
$N=5$ in the energy range of $[11361,13947]$ cm$^{-1}$.
In order to determine the possibility of bifurcations with varying
$N$ in Fig.~\ref{fig1}a,b we show the relevant 
expectation values as a function of
increasing $N$ for the lowest and highest energy states 
respectively in a given polyad.
The expectation values of $\widehat{V}_{sab}$ are relatively small 
and hence not shown in the figures. 
The IPRs in the zeroth order basis and various integrable single resonant
basis are shown in Fig.~\ref{fig1}c,d. For the lowest energy state it is
clear that the $sbb$-Fermi resonance plays an important role. This is
evident from the IPR information as well since the $sbb$ single
resonance basis values are fairly high for the range of $N$ shown.
On the other hand the highest energy states show the importance of
the $saa$-Fermi resonance. The sharp rise in the expectation value
corresponding to $V_{saa}$ between $N=4$ and $N=6$ indicates a change
in the nature of the dynamics which is confirmed by studying the
surface of sections. In both instances the dominance of a single 
$\langle V \rangle$ and the corresponding single resonance basis IPR
establishes the states to be regular and highly localized.

\begin{figure}
\includegraphics*[width=3.5in,height=3.5in]{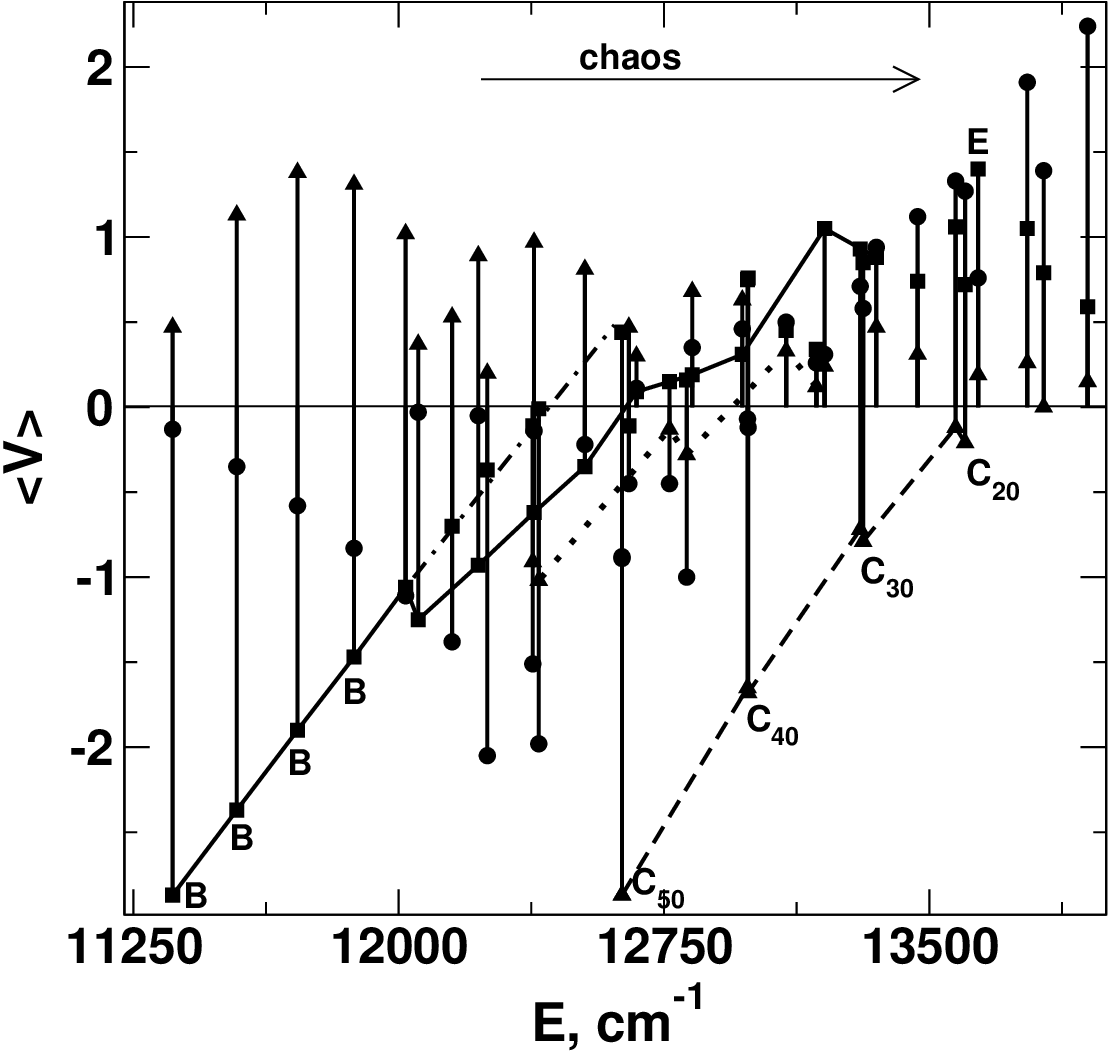}
\caption{Resonance expectation values for the states belonging to
$N=5$ for CHBrClF. The horizontal zero-line is shown for ease
of visualization.
The symbols used are identical to the ones used in
Fig.~\ref{fig1}. The approximate energy at which
large scale chaos sets in the phase space is indicated.
Some of the eigenstate sequences are shown and labeled as in an earlier
work\cite{tay2}.
Note the particularly large values for certain states indicating
localization in state space and phase space. Also note that around
$12020$ cm$^{-1}$ the expectation values $\langle V_{saa} \rangle \approx
\langle V_{sbb} \rangle$ and the splitting of the $'B'$ sequence into
two shown by dash-dotted line.
Further details are provided in the text.}
\label{fig2}
\end{figure}

In Fig.~\ref{fig2} the relevant $\langle V \rangle$ 
are shown for the eigenstates
in $N=5$. Note that a 
large $|V_{k}^{\alpha \alpha}|$ implies a localized state
in the corresponding state space resonance zone and also in the
phase space about a closed orbit. The nature of the closed orbit {\it i.e.,}
stable or unstable is inferred from the 
sign of $V_{k}^{\alpha \alpha}$.
A look at the partial expectation values $V_{k}^{\alpha \alpha}(P_{k})$
will reveal a single dominant contribution leading to the assignment
$(N,P_{k},\nu)$ with $\nu$ denoting an excitation index for states with
the same $P_{k}$. We will not give a list of assignments for the
states in $N=5$ but rather focus on certain 
eigenstate sequences in order to compare
to a recent work\cite{tay1}.
At first glance it is apparent that the $sbb$ resonance
is influencing the lower end of the polyad whereas the $saa$ resonance
is important for the higher energy states. Note, however, that with
increasing energy the $saa$ rapidly gains significance with concomitant
decrease in the $sbb$ influence. 
From Fig.~\ref{fig2} it is seen that around 12020 cm$^{-1}$ 
$\langle V_{saa} \rangle \approx \langle V_{sbb} \rangle$ and the
sequence splits into two branches. This indicates a change in the
underlying phase space and thus localization nature of
the eigenstates on the two branches should be different. This is
supported by examining the phase space and the alternating
assignments provided in a recent work\cite{tay2}.
At the same time towards the middle of
the polyad the Darling-Dennison resonance is playing a key role in
organizing the eigenstates. We plot the $\langle V \rangle$ versus the
energy eigenvalues $E_{\alpha}$ since this immediately reveals the
states that are possibly involved in avoided crossings. 
In Fig.~\ref{fig2} 
certain sequences have been shown
and labeled according to a classification done recently. In this
classification\cite{tay2} $B$ represent nonresonant states, and $C$ represent
states influenced by the Darling-Dennison resonance. From the expectation
values, however, it is not apparent that the first few states are
nonresonant. Indeed our analysis suggests some of these states, for instance
the first three states, are influenced by the $sbb$ Fermi resonance.
That this is indeed the case has been confirmed by the method of
local frequency analysis which shows extensive $sbb$ locking.
In addition the corresponding IPRs in the single $sbb$ resonance
basis are very high suggesting strong influence by the resonance.
Thus state number one is assigned as $(N=5,P_{sbb} \equiv 2v_{s}+v_{b}=10,
v_{a}=0)$ rather than the earlier assignment of $(N=5,v_{a}=0,v_{b}=10)$.

The earlier assignment\cite{tay2}
of Darling-Dennison states essentially agrees with
our results. Certain strongly localized sequence of $C$ states is
immediately revealed by the expectation values as evident
from Fig.~\ref{fig2}. 
It is expected that such states are localized about the stable periodic
orbit associated with the $\widehat{V}_{aabb}$ which has
been confirmed by computing the Husimi distribuion functions.
Our assignment for the state denoted $C_{50}$ is $(N=5,P_{aabb} \equiv
v_{a}+v_{b}=10,v_{s}=0)$ and clearly the approximate polyad
$P_{aabb}$ can be identified with the longitudinal quantum number $n_{l}$
introduced in the earlier work\cite{tay2}. 
Indeed the maximum magnitude of the expectation $V_{aabb}^{\alpha \alpha}$
can be estimated as $P_{aabb}^{2}/4$ and state $C_{50}$ shows close
agreement with this classical estimate. 
The so called transverse
quantum number $n_{t}$ is associated with the degree of excitation
$\nu$ for a given $P_{aabb}$. In the $n_{t}=0$ sequence it is clear from
the figure that with increasing energy other resonances are coming
into effect. The state labeled $C_{20}$, in particular, is strongly
influenced by both the $saa$ and the $sbb$ Fermi resonances and hence
it is inappropriate to classify them as Darling-Dennison states. 
The sequence corresponding to $n_{t}=1$ 
is strongly perturbed
by the $saa$ Fermi resonance. 

Finally state number $16$ and $30$ have been classified in the
previous work as ``chaotic'' states (class $D$) exhibiting
a mixture of class $B$ and class $C$ states. The present analysis
supports the above classification for state $16$, which
incidentally has the lowest IPR among all the states, but clearly
Fig.~\ref{fig2} suggests
state $30$ to be different. 
The large expectation values of $saa$ with
smaller $sbb$ expectation values indicates this state to be
influenced by the $saa$ and $sbb$ resonances. This state can be
nominally assigned as $(N=5,P_{saa} \equiv 2v_{s}+v_{a}=8,v_{b}=2)$.
Similarly state number $33$ was classified as $E(D)$ suggesting
dominant $saa$ character. However state space, 
expectation values (cf. Fig.~\ref{fig2}) and
the various IPRs (integrable single $sbb$ resonance basis IPR is about $0.7$)
point to a state with $sbb$ character. 

\section{Nature of the highly excited bend states in C$_{2}$H$_{2}$}

The pure bending dynamics of acetylene at about $10 000$ cm$^{-1}$ above
ground state is known to be quite complicated due to the strong
interaction between the trans and the cis bending normal modes\cite{jac}. 
However recent work by Jung, Taylor and Jacobson revealed that
the highly excited bending eigenstates were assignable despite
strong chaos in the system\cite{tay3}. 
The key to their dynamical assignments
was the fact that a few periodic orbits were organizing the
dynamics at such high energies and hence influencing the structure
of the eigenstates. Many sequences of eigenstates exhibiting
similar localization patterns about the periodic orbits were identified
leading to the unzipping of the spectrum. 
The purpose of this section is to show that parametric variations can
immediately reveal the existence of such progressions without the
need for the visualization of the phase space and/or the eigenstates.

\begin{figure}
\includegraphics*[width=3.5in,height=3.5in]{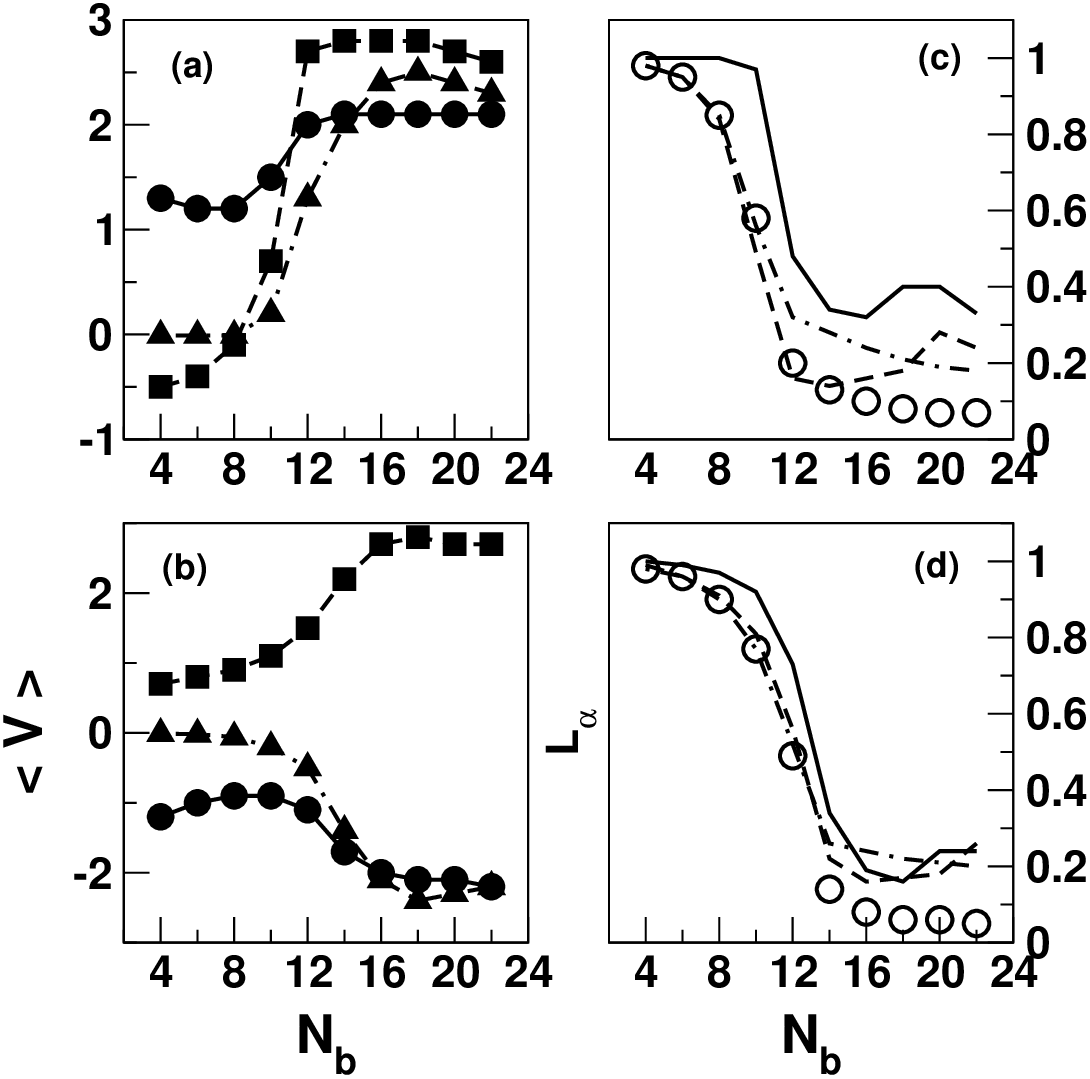}
\caption{Expectation values as a
function of the bend polyad $N_{b}=v_{4}+v_{5}, l=0$
for (a) lowest and (b) highest energy states in a polyad for C$_{2}$H$_{2}$.
Circles, squares and triangles denote the expectation values
$\langle V_{DDI} \rangle$, $\langle V_{DDII} \rangle$, and
$\langle V_{ll} \rangle$ respectively.
Note the sharp change around
$N_{b}=12$ and $N_{b}=14$ for the lowest and highest states respectively.
The IPRs in various basis for the two states are shown in (c) and (d)
as a function of $N_{b}$. IPR in the zeroth order basis are denoted by
open circles and in the various single resonance basis by lines ($H_{0}+
V_{DDI}$ by solid line, $H_{0}+V_{DDII}$ by dashed line, and
$H_{0}+V_{ll}$ by dot-dashed line).}
\label{fig3}
\end{figure}

The effective Hamiltonian appropriate for the study of the bend only
dynamics up to $15 000$ cm$^{-1}$ with an accuracy of $1.5$ cm$^{-1}$
has the form\cite{acetham}
$\widehat{H} = \widehat{H}^{\rm lin} + \widehat{H}^{\rm anh} +
\widehat{H}^{\rm int}$.
The harmonic part $\widehat{H}^{\rm lin} = \omega_{4} \hat{v}_{4}+
\omega_{5} \hat{v}_{5}$ and the anharmonic part is given by
\begin{equation}
\widehat{H}^{\rm anh}=\sum_{j,k=4,5}(x_{jk} v_{j} v_{k}+g_{jk} l_{j} l_{k})
+ \sum_{jkl=4,5} y_{jkl} v_{j} v_{k} v_{l}
\end{equation}
with $v_{4}$ and $v_{5}$ representing the number of quanta in the trans
and cis normal modes respectively. The degenerate bends further
require the vibrational angular momenta $l_{4}$ and $l_{5}$. Thus the
zeroth order states, eigenstates of $\widehat{H}^{\rm lin} + 
\widehat{H}^{\rm anh}$, are denoted by $|v_{4}^{l_{4}} v_{5}^{l_{5}} \rangle$.
However $\widehat{H}^{\rm int}=\widehat{V}_{DDI}+\widehat{V}_{DDII}+
\widehat{V}_{ll}$ couples the zeroth order states via the
off-diagonal anharmonic resonances. 
$\widehat{V}_{DDI}$ is
a Darling-Dennison resonance leading to exchange of quanta between
the two modes at constant $l_{4}$ and $l_{5}$ whereas
$\widehat{V}_{ll}$ is a vibrational $l$-resonance which leads to
exchange of vibrational angular momentum only between the two modes.
$\widehat{V}_{DDII}$ results in both exchange of quanta and vibrational
angular momentum between the two modes.
For the fitted parameter values and form of the perturbation operators
we refer the reader to the previous works\cite{acetham,tay3}.
Due to the nature of the resonant couplings there are two conserved
quantum numbers or polyads denoted by $N_{b}=v_{4}+v_{5}$ and $l=l_{4}+l_{5}$.
In addition eigenstates respect certain symmetries and hence labelled
by $\pm$ (parity under say $l_{4} \rightarrow -l_{4}$) 
and $g/u$ ($v_{5}$ even/odd). There are
a total of $81$ states for the polyad $N_{b}=16$ 
spanning an energy range of $[10239,11255]$ cm$^{-1}$.
Although it is possible to analyze all of the eigenstates, in this letter
we focus on the subset
of eigenstates with $N_{b}=16$ and $l=0$ and symmetry class $g+$ {\it i.e.,}
states belonging to $[16,0]^{g+}$. Note that there are a total of
$25$ states in $[16,0]^{g+}$. As in the previous section we will be 
interested in the expectation values (three of them for acetylene)
$V^{\alpha \alpha}_{j}$. 

To begin with we consider, in analogy with the previous section,
the possibility of various
bifurcations occuring in C$_{2}$H$_{2}$ with varying polyad $N_{b}$.
Earlier studies\cite{jac} have revealed that the 
lowest (trans bend) and highest
(cis bend) energy eigenstates in a given polyad undergo a sharp
change in character with increasing $N_{b}$. More precisely, around
$N_{b}=12$ the lowest state is not the usual trans bend and becomes
a local bend whilst around $N_{b}=14$ the highest state changes
character from a cis bend to the so called counter-rotator mode.
Such changes are characterisitc of bifurcations which lead to
new types of modes (dynamics) and we expect the expectation values to
be sensitive indicators of these changes. In order to show this in
Fig.~\ref{fig3}a,b we show the $\langle V \rangle$ for the lowest and the
highest energy states as a function of $N_{b}$. As expected the
$\langle V \rangle$ very clearly indicate the sharp nature of the
bifurcations in agreement with the previous observations. In 
Fig.~\ref{fig3}c,d we have also shown 
the IPRs of the states in various basis for
comparison. Note that the IPRs also indicate the transition. However,
the various IPRs in the post transition regime take on very small
values and by definition this indicates that many zeroth order normal
modes are contributing to the eigenstates and thus
suggesting highly mixed states. 
Contrast this with the fact that 
all of the
expectations are very large suggesting highly localized states.  
The resolution to these apparently conflicting observations lies in
realizing that new types of dynamics {\it i.e.,} periodic orbits
created due to the bifurcations
are influencing the eigenstates and the zeroth order normal modes
are a poor basis to understand the new modes.
It is important to note that IPR in any single resonance (dressed) basis
is also an insufficient indicator of the new modes. The expectation values
on the other hand are quite sensitive and the nature of the new
modes can be deciphered with further analysis\cite{tay1}.

\begin{figure}
\includegraphics*[width=3.5in,height=3.5in]{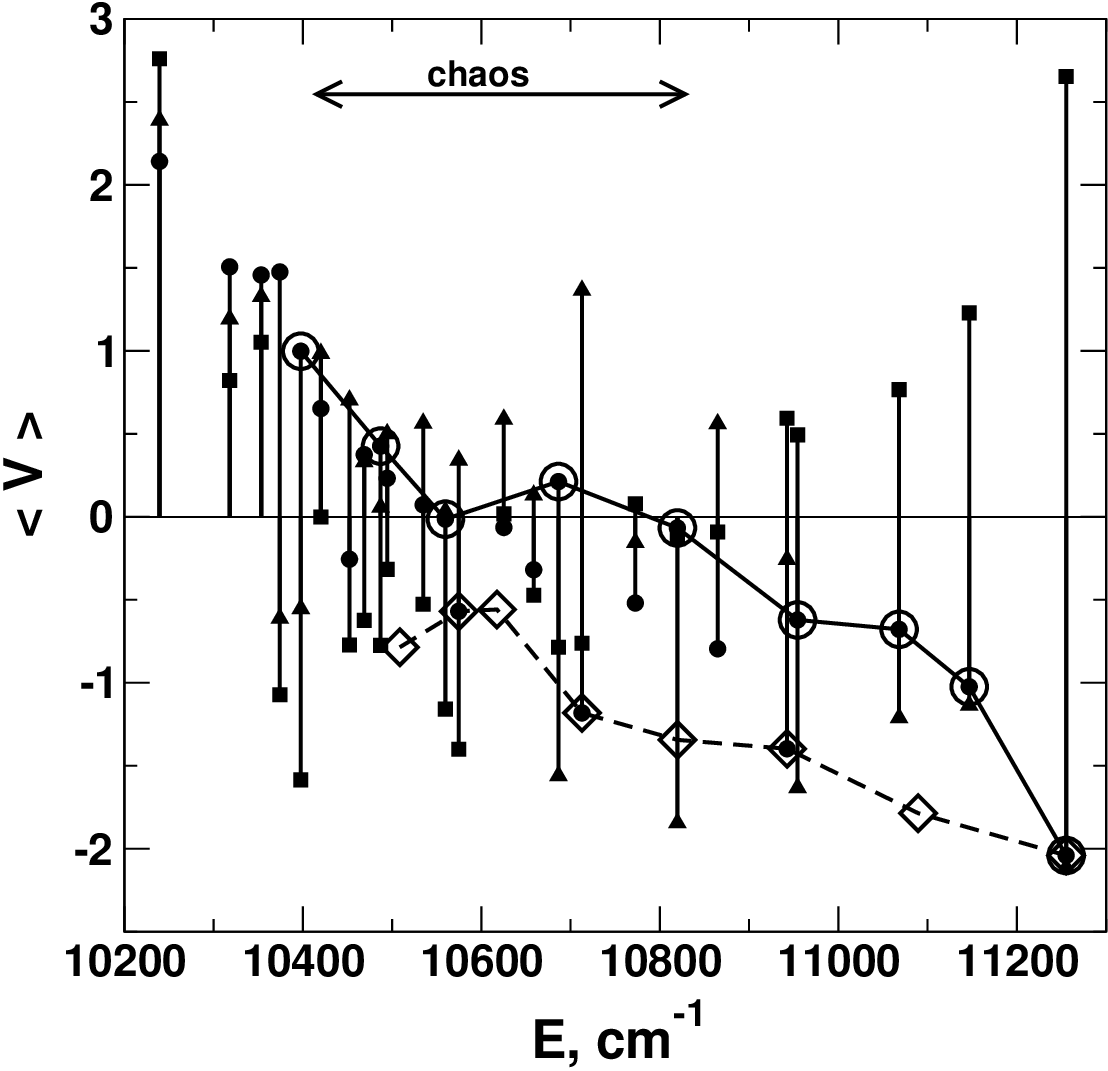}
\caption{Expectation values for the states belonging to the $[16,0]^{g+}$
polyad of C$_{2}$H$_{2}$.
Symbols used are identical to those in Fig.~\ref{fig3}.
Open Diamonds mark a sequence of eigenstates representing
the family I and open circles indicate the sequence of eigenstates
representing the family II as in an earlier work\cite{tay3}.
The lines are drawn as a guide
to the eye. Note some of the family I states are $|u\rangle^{+}$ states
shown in this plot for clarity. The approximate energy region over which
considerable chaos exists in the classical dynamics is also indicated.
See text for discussions.}
\label{fig4}
\end{figure}

We now turn to an analysis of the states in $[16,0]^{g+}$.
In Fig.~\ref{fig4} the expectation values 
are shown for all the states. Based on
previous observations it is easy to identify the local bend and the
counter rotating states at the energetic minimum and maximum of the polyad.
The appearance of rather complicated behaviour in the middle of the
polyad is related to the occurence of bifurcations with varying
energy\cite{tay3}. Further understanding can be gained by analyzing the
reduced classical Hamiltonian\cite{tay3}:
\begin{eqnarray}
H({\bf J},&{\bm \psi}&;{\bf K}) = 
H_{0}({\bf J};{\bf K}) + 
2 s_{45} f_{a}({\bf J};{\bf K}) \cos \psi_{a} \nonumber \\
&+& 2 r_{45} f_{b}({\bf J};{\bf K}) \cos \psi_{b}\\ 
&+& 2 t_{45} [f_{-}({\bf J};{\bf K}) \cos (\psi_{a}-\psi_{b}) +
f_{+}({\bf J};{\bf K}) \cos (\psi_{a}+\psi_{b})]\nonumber
\end{eqnarray}
where ${\bf J} = (J_{a},J_{b}) \equiv ((v_{4}-v_{5})/4,(l_{4}-l_{5})/4)$, 
${\bf K} = (K_{a},K_{b}) \equiv ((N_{b}+2)/4,l/4)$, and
$t_{45} = (r_{45}+2 g_{45})/4$. The angles $\psi_{a}$ and $\psi_{b}$
correspond to the DDI and the vibrational-$l$ resonances respectively.
The angle combinations $\psi_{a} \pm \psi_{b}$ correspond to
the DDII resonance. The fixed points of the above Hamiltonian {\it i.e.,}
$(\dot{\bf J},\dot{\bm \psi})=({\bf 0},{\bf 0})$ represent the
periodic orbits of the full system. It is easy to see that in the
angle space the fixed points are $\psi_{a,b}=0,\pm \pi$.
From our discussions we expect the functions $f_{a},f_{b}$, and
$(f_{-},f_{+})$ at the fixed points to indicate the 'dynamical' nature
of a specific eigenstate. In particular, the signs of the various
expectations provide information on the nature of eigenstate
localization in the $(\psi_{a},\psi_{b})$ space. It is easy to see that
four possible sign combinations $+++,+--,--+$, and $-+-$ are possible
corresponding to the localization about $(0,0),(0,\pi),(\pi,\pi)$, and
$(\pi,0)$ respectively. At this stage we emphasize that highly localized
eigenstates imply a specific sign combination and large magnitude
expectation(s). 
Thus, for instance, from Fig.~\ref{fig4} we anticipate that the
states $|g \rangle_{1}^{+},|g \rangle_{4}^{+},|g \rangle_{25}^{+}$,
and $|g \rangle_{17}^{+}$ are highly localized states about
$(0,0),(0,\pi),(\pi,\pi)$, and $(\pi,0)$ respectively. 
These states exhibit localization in phase space as well. 
We find a total of $18$ states that belong to
one of the four sign combinations. On the other hand there
are states that do not belong to one of the above four classes.
For example $|g \rangle_{14}^{+}$ has the sign combination $-++$
and small expectation values. This suggests a fairly delocalized
state which is hard to assign dynamically. Similarly 
state $|g \rangle_{19}^{+}$ comes with the sign combination $---$
but Fig.~\ref{fig4} clearly
implies a localized state influenced by the vibrational-$l$ resonance
alone. Interestingly such states are precisely the ones that
were multiply assigned in the previous study\cite{tay3}.

Jung {\it et. al.} have used the semiclassical representation of
the eigenstates\cite{sibmc} 
in the $(\psi_{a},\psi_{b})$ space to provide
a sequence of similarly localized states\cite{tay3}. 
Two main families, I and II,
were identified based on the excitation along specific periodic orbits.
In Fig.~\ref{fig4} we show two such sequences, one from each family,
as seen from the perspective of the expectation values. It is
apparent that the family I is more robust than the family II sequence.
This is directly related to the fact that the periodic orbit 
underlying family I undergoes far fewer bifurcations as compared
to the periodic orbit corresponding to family II. However even in
the family I sequence two states $|u \rangle_{12}^{+}$ 
and $|g \rangle_{13}^{+}$ are significantly perturbed.
There can be many sources for such perturbations and in this case it happens
to be an avoided crossing. Nevertheless from Fig.~\ref{fig4} it is
clear that the state $|g \rangle_{13}^{+}$ is localized mainly due
to the DDII resonance.

\section{Conclusions}
In this letter we have demonstrated the utility of eigenstate expectation
values of resonant perturbations in understanding the nature of highly
excited vibrational states. A particularly large expectation value
implies that the associated eigenstate is localized and influenced by
a closed orbit in the classical phase space. 
Existence of localized states and their sequences can be ascertained
by inspecting the expectation values (magnitude and sign) and
the IPRs irrespective of wether the classical phase space
is (near)-integrable, mixed or chaotic. 
In case of complete chaos, from random matrix
theory arguments, a typical expectation value is expected\cite{haake}
to be $1/\sqrt{N_{s}}$.
The examples studied here are quite far from such a limit.
Combined
with the relative signs of the
expectation values it is possible to identify 
specific eigenstate sequences. 
Sensitivity of the expectation values to the bifurcations provides
information on the birth of new types of modes and the resulting
perturbations on the eigenstate sequences. Currently it is not possible
to identify the type of bifurcation that occurs and this aspect needs
further study. It is important to note that we are not advocating
the use of periodic orbits to compute the expectation values which
is a difficult task. Instead we are emphasizing the 
manifestation and utility of
such classical structures in a fundamental quantum object {\it i.e.,}
an expectation value. 
Similar philosophy has been adopted in the 'vibrogram' or $(E,\tau)$
approaches to study resonant systems\cite{greg,gasp} 
This classical-quantum correspondence aspect
of the expectation values can be easily applied to multidimensional
coupled systems without the need for determining/visualization of
phase space, periodic orbits or eigenstates. 
Indeed preliminary work on a coupled 4-mode
effective Hamiltonian suggests that such an approach is useful.

Finally we note that Jacobson and Field have recently\cite{jfield} studied
expectation values of resonance operators for a nonstationary state.
Choosing the nonstationary state 
to be a zeroth order bright state $|{\bf v}\rangle$
it was shown that the time-dependent expectations 
$\langle {\bf v}(t)|\widehat{V}_{k}|{\bf v}(t)\rangle$
indicate the resonances important for dynamics over a particular time
interval. Relation to the present work is realized by the fact that
\begin{equation}
\lim_{T \rightarrow \infty} \frac{1}{T} \int_{0}^{T} 
\langle {\bf v}(t)|\widehat{V}_{k}|{\bf v}(t)\rangle = 
\sum_{\alpha} |\langle {\bf v}|\alpha \rangle|^{2} V_{k}^{\alpha \alpha}
\end{equation}
Thus the time-dependent expectations are, in the infinite time limit,
nothing but intensity weighted eigenstate expectation values. Such
an object, the intensity-velocity correlator, has been recently
introduced and studied in great detail\cite{nickjcp}.

\section{Acknowledgements}
This work is supported by funds from the Department of Science and
Technology, India.


\begin{thebibliography}{}
\bibitem{rev1}{R. Marquardt and M. Quack, {\em Encyclopedia of
Chemical Physics and Physical Chemistry}, Vol.I, Ed. J. H. Moore,
IOP, Bristol, 2001. }
\bibitem{rev2}{T. Uzer, Phys. Rep. {\bf 199}, 73 (1991).}
\bibitem{rev3}{D. J. Nesbitt, R. W. Field, J. Phys. Chem. {\bf 100},
12735 (1996).}
\bibitem{rev4}{M. Gruebele, Adv. Chem. Phys. {\bf 114}, 193 (2000).}
\bibitem{cvpt1}{A. B. McCoy, E. L. Sibert III in {\em Dynamics of
Molecules and Chemical reactions}, Ed. R. E. Wyatt and J. Z. H. Zhang,
Dekker, NY 1996.}
\bibitem{cvpt2}{M. Joyeux, D. Sugny, Can. J. Phys. {\bf 80}, 1459 (2002).}
\bibitem{aav}{A. J. Lichtenberg, M. A. Lieberman, {\em Regular and
Stochastic Motion}, Springer, NY 1983.}
\bibitem{leto}{See, M. V. Kuzmin, A. A. Stuchebrukhov in {\em Laser
Spectroscopy of Highly Vibrationally Excited Molecules}, Ed. V. S. Letokhov,
pp. 178, Adam Hilger, Bristol, 1989.}
\bibitem{kell}{M. E. Kellman, Adv. Chem. Phys. {\bf 101}, 590 (1997).}
\bibitem{joy}{M. Joyeux, S. C. Farantos, R. Schinke, J. Phys. Chem. A,
{\bf 106}, 5407 (2002).}
\bibitem{jac}{M. P. Jacobson, R. W. Field, J. Phys. Chem. A {\bf 104},
3073 (2000).}
\bibitem{tay1}{M. P. Jacobson, C. Jung, H. S. Taylor, R. W. Field,
J. Chem. Phys. {\bf 111}, 600 (1999).}
\bibitem{tay2}{C. Jung, E. Ziemniak, H. S. Taylor, J. Chem. Phys. {\bf 115},
2499 (2001).}
\bibitem{field}{H. Ishikawa, R. W. Field, S. C. Farantos, M. Joyeux, J. Koput,
C. Beck, R. Schinke, Annu. Rev. Phys. Chem. {\bf 50}, 443 (1999).}
\bibitem{ksgse1}{S. Keshavamurthy, G. S. Ezra, J. Chem. Phys. {\bf 107},
156 (1997).}
\bibitem{kspccp}{A. Semparithi, S. Keshavamurthy, Phys. Chem. Chem. Phys.
{\bf 5}, 5051 (2003).}
\bibitem{commcdbr}{However, see C. Jung, C. Mejia-Monasterio, H. S. Taylor,
J. Chem. Phys. {\bf 120}, 4194 (2004).}
\bibitem{ram}{R. Ramaswamy, R. A. Marcus, J. Chem. Phys. {\bf 74},
1379 (1981).}
\bibitem{rose}{J. P. Rose, M. E. Kellman, J. Chem. Phys. {\bf 104},
10471 (2000).}
\bibitem{wu}{P. Wang, G. Wu, Chem. Phys. Lett. {\bf 371}, 238 (2003).}
\bibitem{kay}{B. Ramachandran, K. G. Kay, J. Chem. Phys. {\bf 99},
3659 (1993).}
\bibitem{ksjpca}{S. Keshavamurthy, J. Phys. Chem. A {\bf 105}, 2668 (2001).}
\bibitem{child}{M. S. Child, J. Mol. Spec. {\bf 210}, 157 (2001).}
\bibitem{weissjort}{Y. Weissman, J. Jortner, J. Chem. Phys. {\bf 77},
1486 (1982).}
\bibitem{ksjcp}{A. Semparithi, V. Charulatha, S. Keshavamurthy,
J. Chem. Phys. {\bf 118}, 1146 (2003).}
\bibitem{eck}{B. Eckhardt, S. Fishman, K. M\"{u}ller, D. Wintgen,
Phys. Rev. A {\bf 45}, 3531 (1992).}
\bibitem{gutz}{M. C. Gutzwiller, {\em Chaos in Classical and Quantum
Mechanics}, Springer, NY 1990.}
\bibitem{beil}{A. Beil, D. Luckhaus, M. Quack, Ber. Bunsenges. Phys. Chem.
{\bf 100}, 1853 (1996).}
\bibitem{tay3}{C. Jung, H. S. Taylor, M. P. Jacobson, 
J. Phys. Chem. A {\bf 105}, 681 (2001).}
\bibitem{acetham}{M. P. Jacobson, J. P. O'Brien, R. J. Silbey, R. W. Field,
J. Chem. Phys. {\bf 109}, 121 (1998).}
\bibitem{sibmc}{E. L. Sibert, A. B. McCoy, J. Chem. Phys. {\bf 105},
469 (1996).}
\bibitem{haake}{F. Haake, {\em Quantum Signatures of Chaos}, 2$^{nd}$ edition,
Springer, Berlin, 2001. Considerable work has been done towards
understanding the parametric variation in a variety of systems and chapter $6$
of this book provides an introduction to this vast field.}
\bibitem{greg}{G. S. Ezra, Adv. Class. Traj. Meth. {\bf 3}, 35 (1998)
and references therein.}
\bibitem{gasp}{P. Gaspard, I. Burghardt, Adv. Chem. Phys. {\bf 101},
491 (1997).}
\bibitem{jfield}{M. P. Jacobson, R. W. Field, Chem. Phys. Lett. {\bf 320},
553 (2000).}
\bibitem{nickjcp}{See S. Keshavamurthy, N. R. Cerruti, S. Tomsovic,
J. Chem. Phys. {\bf 117}, 4168 (2002) and references therein.}
\end{thebibliography}
\end{document}